\documentclass[conference]{IEEEtran}
\IEEEoverridecommandlockouts
\usepackage{cite}
\usepackage{amsmath,amssymb,amsfonts}
\usepackage{algorithm}
\usepackage{algpseudocode}
\usepackage{color}
\usepackage{graphicx}
\usepackage{textcomp}
\graphicspath{{../figures/}}

\def\BibTeX{{\rm B\kern-.05em{\sc i\kern-.025em b}\kern-.08em
    T\kern-.1667em\lower.7ex\hbox{E}\kern-.125emX}}

\def\E#1{\ensuremath{\text{E}\left \{#1 \right \}}}

\begin{document}

\title{A Software-based Low-Jitter Servo Clock for Inexpensive Phasor Measurement Units }

\author{
\IEEEauthorblockN{Pietro Tosato, David Macii, \\ Daniele Fontanelli and Davide Brunelli}
\IEEEauthorblockA{
University of Trento\\
Trento, Italy
}
\and
\IEEEauthorblockN{David Laverty} 
\IEEEauthorblockA{
Queen's University of Belfast\\
Belfast, U.K. \\
}

}

\maketitle

\begin{abstract}
This paper presents the design and the implementation of a servo-clock (SC)
for low-cost Phasor Measurement Units (PMUs). The SC relies on a classic Proportional Integral (PI) controller, which has been properly tuned to minimize the synchronization error due to the local oscillator triggering the on-board timer. The SC
has been implemented into a PMU prototype developed within the OpenPMU project using a BeagleBone Black (BBB) board. The distinctive feature of the proposed solution is its ability to track an input Pulse-Per-Second (PPS) reference with good long-term stability and with no need for specific on-board synchronization circuitry. Indeed, the SC implementation relies only on one co-processor for real-time application and requires just an input PPS signal that could be distributed from a single substation clock.
\end{abstract}

\begin{IEEEkeywords}
Servo clock, time synchronization, syntonization, Phasor Measurement Units, smart grid. 
\end{IEEEkeywords}

\section{Introduction}
\label{sec:Intro}

Time synchronization is a crucial part of every distributed measurement system. Synchrophasor measurement in transmission and distribution systems is a well-known
application field in which time synchronization plays a key role, e.g. for
power system state estimation~\cite{Ree10}, topology detection~\cite{Cavraro15} or loss-of-mains protection~\cite{Laverty15}. 
Phasor Measurement Units (PMU) are complex instruments, requiring usually high-end hardware that perform timestamped measurements of voltage and current amplitude, phase, frequency and rate of change of frequency (ROCOF) synchronized to the Coordinated Universal Time (UTC). From a functional point of view, PMUs can be decomposed into component parts: an acquisition subsystem, a synchronization module, 
and finally a digital signal processing subsystem.
Several scientific contributions about the impact of PMU synchronization uncertainty in power systems have already been  proposed in the literature~\cite{Bazerque2016}. 
For instance, analytical methods for mitigating the effect of time synchronization on the grid state estimation are presented in~\cite{Todescato2017, Yang2013}. 
Some insight from a real network can be found in the work by Della Giustina et al.~\cite{Giustina2014}, who analyze the timing requirements for power quality measurements. 
The IEEE Standard C37.242-2013 on PMU synchronization, calibration, testing, and installation highlights that the main contributors to estimation uncertainty can be identified in (i) synchronization issues, (ii) noise and distortion in the input channel and acquisition circuitry, (iii) intrinsic accuracy limits of the adopted digital signal processing algorithm and, finally, (iv) possible communication problems~\cite{c37.242-2013}. 
In general, no information on the weight of these contributions on overall PMU accuracy is available. The IEEE Standard C37.118.1-2011 and its Amendment IEEE  C37.118.1a-2014 express the overall synchrophasor measurement accuracy with a single parameter, namely the Total Vector Error (TVE), which depends on both amplitude and phase estimation uncertainties. While both documents prescribe various TVE boundaries in different testing conditions~\cite{c37.118-2011,c37.118-2014}, no specific limits for time synchronization uncertainty or jitter are explicitly reported. Under the overoptimistic assumption that the amplitude measurement uncertainty is negligible, time errors within $\pm31$ $\mu$s or $\pm26$ $\mu$s for 50 Hz or 60 Hz systems, respectively, are small enough to keep phase estimation accuracy below 10 mrad and, consequently, TVE~$\leq1$\%, which is the strictest limit reported in~\cite{c37.118-2011,c37.118-2014}. However, since amplitude and phase estimation uncertainties are usually both significant in practice, ``a time source that reliably provides time, frequency, and frequency stability at least 10 times better than the values above is highly recommended"~\cite{c37.118-2011}. Thus, as a rule of thumb, synchronization accuracy within $\pm1$ $\mu$s is currently considered to be adequate in most power systems applications, as the corresponding maximum phase errors (in the order $\pm0.4$ mrad) are usually much smaller than those due to other uncertainty contributions. 
However, the evolution of smart active distribution grids as well as the emerging need to measure phasor angle differences smaller than 1 mrad (e.g. over short lines) could demand more accurate synchrophasor measurements than those possible  nowadays~\cite{Borghetti11, Wen15, Barchi2015}. As a result, tighter synchronization accuracy might be needed in the future. 
%
It is worth emphasizing that PMUs require not only {\em synchronization} (i.e. time offset compensation with respect to UTC) to properly timestamp measurement data, but also {\em syntonization} (i.e. clock rate adjustment) to enable coherent sampling of voltage or current waveforms in ideal conditions.
Commercial PMUs generally include specific hardware modules for time synchronization, most notably GPS receivers or IRIG-B (Inter-Range Instrumentation Group time codes) decoders, which are supposed to be used to discipline the sampling clock as well, e.g. through some hardware Phase Lock Loop (PLL) or other more sophisticated custom techniques.
For instance, in~\cite{Yao2018} Yao et al. describe a way to compensate 
for the sampling time errors caused by the division
remainder between the desirable sampling rate and the oscillator
frequency. 
An alternative approach to achieve both {\em synchronization} and {\em syntonization} is through Servo Clocks (SC), e.g. based on Proportional Integral (PI) controllers~\cite{Exel2013, Eidson2006c}. 
The most common examples of SCs are those developed for Ordinary and Boundary Clocks of IEEE~1588 devices~\cite{Correll05, Machizawa2008, Ferencz2013}. In general, there are just a few comprehensive analyses of SCs. One of them, is provided by Chen et al.~\cite{Chen2017} who propose an optimized SC for distributed motion control systems based on EtherCAT. However, the design and implementation of SCs for PMUs is a topic seldom covered in the scientific literature. Even the recently released IEEE Standard C37.238-2017 dealing with a profile of the IEEE~1588 Precision Time Protocol (PTP) for power systems application does not report any indication about SC design~\cite{c37.238-2017}. 
This research work is part of the `OpenPMU' project\footnote{http://www.OpenPMU.org}, an international project 
whose purpose is to develop a fully open-source PMU for power system analysis and research~\cite{Laverty2013}.
In particular, this paper deals with a SC for the `OpenPMU' platform described in~\cite{Zhao2017}. The SC has been 
 designed and optimized to minimize the synchronization errors due to the local crystal oscillator (XO)  and generates the signal to sample the input waveform as well.
 The main advantage of the proposed solution is that the SC relies only on a Programmable Real-Time Unit (PRU) available in the embedded platform, with no need for specific synchronization hardware except for an external Pulse Per Second (PPS) reference signal, which could be provided by a common GPS receiver (or substation clock) and shared among multiple PMUs. 
The rest of the paper is structured as follows. First, in Section~\ref{sec:implementation}, the resources
available to implement the SC for the `OpenPMU' platform are described in brief. Then, in Section~\ref{sec:model}, a mathematical model of the SC is defined and the related design criteria are explained. Finally, in Section~\ref{sec:results} the results of various experiments showing SC performance are reported. Section~\ref{sec:conclusion} concludes the paper and outlines future work.

\section{Servo Clock Architecture}
\label{sec:implementation}

\begin{figure}[t]
	\centering
	\includegraphics[page=7, trim={2cm 3.5cm 2cm 3.5cm}, clip, width=\columnwidth]{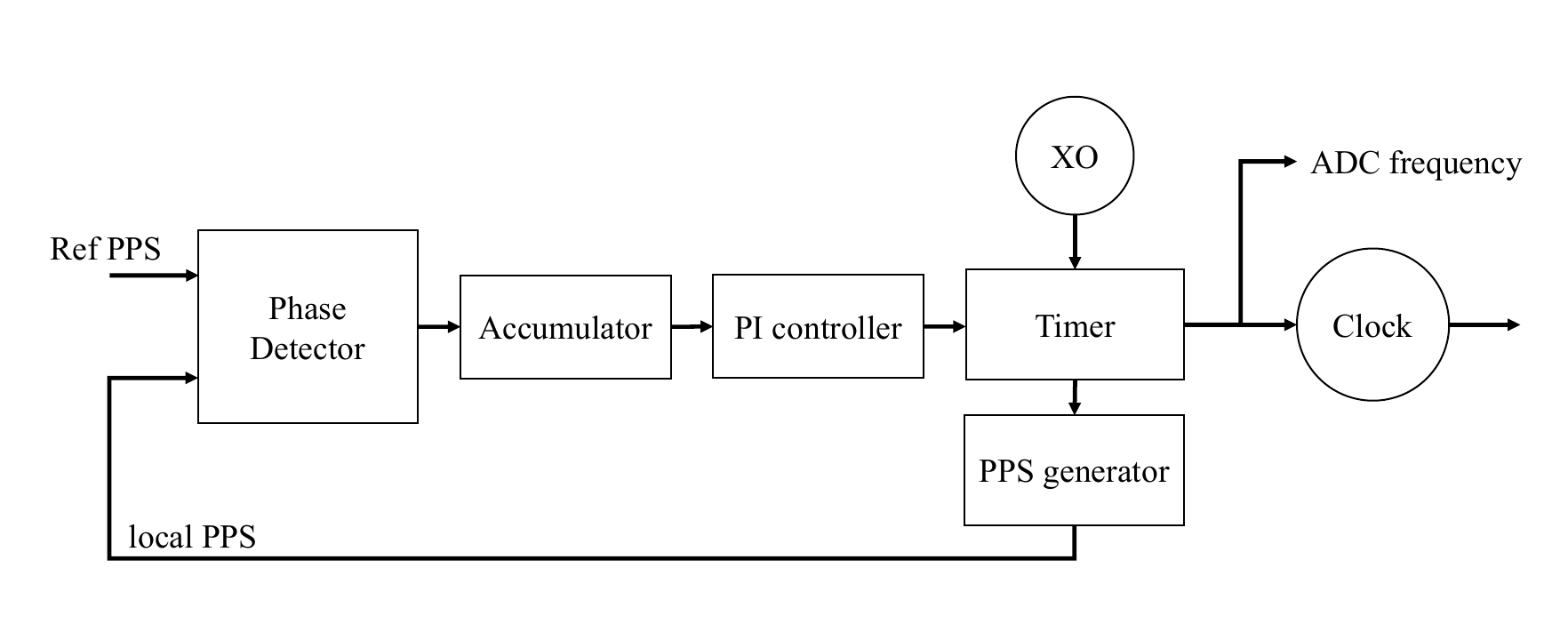}
	\caption{Architecture of the SC implemented on the PRU of the BBB.}
	\label{fig:implementation}
\end{figure}

Unlike typical PMU implementations, the acquisition stage of the `OpenPMU' platform described in~\cite{Zhao2017} is fairly simple and relies on a Beaglebone Black (BBB) board. 
This is a low-cost commercial embedded system, which has been recently used in a variety of projects, including I/O signal synchronization~\cite{Alanwar2017} and
PMU algorithm prototyping~\cite{Tosato2018}.
A distinctive feature of the BBB is its Sitara AM3358 microprocessor, that include a 1-GHz ARM Cortex-A8 microprocessor running a Linux kernel, 
and two 200-MHz co-processors, called Programmable Real-time Units
(PRU). The PRUs can be used to perform specific real-time tasks, since
they can be programmed at a low-level, i.e. without using any
operating system. The PRUs are provided with a rich set of peripherals
(including timers), besides direct access to
General-Purpose Input-Output (GPIO) pins. On the other hand, the PRU
computational capabilities are limited: no Floating-Point Unit is
present, local memory is quite small (only 8~KB plus a 12~KB of shared
DRAM) and asynchronous interrupt handling is not possible.
Despite such limitations, the basic idea of the solution proposed in this paper is to use one of the PRUs to fully implement a SC running in parallel to the main ARM core
in order to  {\em synchronize} and {\em syntonize} the data acquisition system described in~\cite{Zhao2017}. 
The architecture of the proposed SC is shown in
Fig.~\ref{fig:implementation}. A purely software indirect frequency
synthesizer is used to generate the 12.8-kHz signal clocking the
Analog-to-Digital Converter (ADC) of the acquisition stage.  Since a
real 12.8-kHz Voltage-Controlled Oscillator (VCO) is not available on
the BBB, this is emulated by means of one of the timers of the PRU,
clocked at 200~MHz and configured to be reloaded automatically. While
the nominal timeout to be loaded into the timer is 15625 ticks, its
actual value changes as a function of the corrective action performed
by the internal controller. The 12.8~kHz signal is used to increment
both the system clock (which is implemented as a software counter
properly initialized with a UTC timestamp as soon as it is available)
and a second counter (labeled as {\em PPS generator} in
Fig.~\ref{fig:implementation}) that generates a PPS signal. The
difference in time (measured with a resolution of 5~ns) between the
external (i.e. reference) PPS signal and the local one is integrated
by a digital accumulator in order to compute the time error, which
finally drives a Proportional-Integral (PI) controller adjusting the
clock rate, as customary in SC design.  The equivalence between the
system considered and a classic SC is demonstrated in Section~\ref{sec:model}.

\section{Model Description and PI Controller design} 
\label{sec:model}

\begin{figure}[t]
	\centering 
	\begin{tabular}{c}
		\includegraphics[page=3, trim={3.3cm 5.5cm 3cm 5.5cm}, clip, width=\columnwidth]{figures/models} \\
		(a) \\
		\includegraphics[page=1, trim={6cm 6.5cm 6cm 6.7cm}, clip, width=0.85\columnwidth]{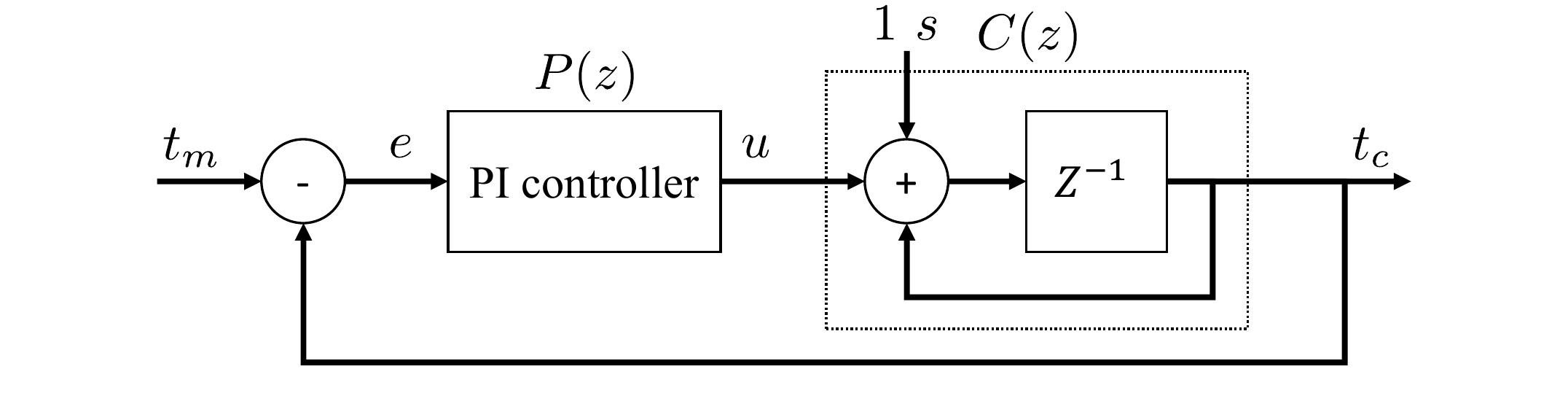} \\
		(b)
	\end{tabular}
	\caption{Equivalent models of the implemented SC before (a) and after (b) reduction.}
	\label{fig:model}
\end{figure}

The SC shown in Fig.~\ref{fig:implementation} can be modeled as a
discrete-time linear system discretized at 1 Hz, as shown in in Fig.~\ref{fig:model}(a). 
Let $\tau_m(k)$ and $\tau_c(k)$ be the periods of the PPS signals
at the input and at the output of the SC, respectively, at the $k-$th sampling second.
Since both signals are affected by phase and frequency noises
and by a relative frequency offset, the time synchronization error $e(k)$
between them is simply given by:
\begin{equation}
\label{eq:error1}
e(k+1) = e(k) + \tau_m(k) - \tau_c(k) ,
\end{equation}
Therefore, the equivalent model depicted in Fig.~\ref{fig:model}(a) 
can be easily reduced to the
classic SC model shown in Fig.~\ref{fig:model}(b), where
symbols $t_m$ and $t_c$ denote the reference time and the time measured by the SC, respectively. 
Based on this simplified model, 
the SC in the $z$-domain basically consists of two subsystems, i.e. 
the clock itself, whose transfer function is 
\begin{equation}
C(z) = \frac{1}{z-1} , 
\label{eq:clock}
\end{equation}
and the PI controller, which can be obtained from the classic backward Euler integration method, i.e.
\begin{equation}
P(z) = K_P + K_I \frac{z}{z-1}, 
\label{eq:control}
\end{equation}
where $K_P$ and $K_I$ are the proportional and integral gains,
respectively.  Of course, SC stability depends on the position of the poles of the 
closed-loop transfer function
\begin{equation}
H(z)\!=\!\frac{P(z)C(z)}{1+P(z)C(z)} \!=\!  \frac{(z-1)K_P + zK_I}{(z-1)^2 \!+\! (z-1)K_P + zK_I}.
\label{eq:cltf}
\end{equation}
Moreover, coefficients $K_P$ and $K_I$ in~\eqref{eq:control} should be
tuned in order to meet given performance requirements, in terms of
convergence time or output uncertainty.  To this
end,~\eqref{eq:clock}, \eqref{eq:control} and~\eqref{eq:cltf} can be
expressed using difference equations.  Nevertheless, the control
design formulation is slightly complicated by the fact that i) the
resolution of the SC is $1/f_{XO}$ (with $f_{XO} = 200$~MHz), and ii)
the nominal frequency of the emulated VCO is $f_0 = 12.8$~kHz.
Therefore, all time quantities (as well as the controller output) have
to be expressed in ticks of an ideal 200~MHz oscillator rather than in
seconds.  Thus, if $n_t=f_{XO}/f_0$ denotes the number of ticks in one
nominal period,
%
%
%
%
%
then the dynamic of the frequency of the disciplined emulated VCO is given by
\[
f_c(k+1) = f_0 (1 + \alpha) + b \eta(k) + b u(k) ,
\]
%
%
where $b = 1/n_t$, $\alpha$ is the
relative frequency offset of the SC local oscillator with respect to the input reference signal when the SC is in open loop, 
$\eta(k)$ represents the jitter of
the SC accumulated during one second and, finally, $u(k)$ is the control
action. 

For what concerns the error, it follows immediately from~\eqref{eq:error1} that
\[
e_q(k+1) = e_q(k) + \tau_{m_q}(k) - \tau_{c_q}(k) ,
\]
where $e_q(k)$ is the time error $e(k)$ expressed in ticks,
\[
\tau_{m_q}(k)=n_tf_0 + \nu(k) ,
\]
is the period of the PPS reference input signal expressed in ticks,
$\nu(k)$ is the jitter of the input PPS signal and
\[
\tau_{c_q}(k)=n_tf_c(k) ,
\]
is the period of the PPS output of the SC, again expressed in
ticks. The control action dynamics is instead given by
\[
u(k+1) = u(k) + K_P(e_q(k+1) - e_q(k)) + K_I e_q(k+1) .
\]
If vector $q(k) = [e_q(k), f_c(k), u(k)]^T$ (with 
$q(0) = [0, f_0, 0]^T$) denotes the
aggregated state of the system in closed loop,  the
previous equations we can be rewritten more compactly as
\[
q(k+1) = A q(k) + G \tau_{m_q}(k) + C \eta(k) + F f_0,
\]
where
\[
A = \begin{bmatrix}
1 & -n_t & 0 \\
b K_I & -(K_P + K_I) & b \\
K_I & -n_t (K_P + K_I) & 1
\end{bmatrix},\,\,\,
G = \begin{bmatrix}
1 \\
b (K_P + K_I) \\
K_P + K_I
\end{bmatrix} ,
\]
\[
C = \begin{bmatrix}
0 \\
b \\
0
\end{bmatrix}\,\,\, \mbox{and}\,\,\,
F = \begin{bmatrix}
0 \\
(1 + \alpha) \\
0
\end{bmatrix} .
\]
\begin{figure}[t]
	\centering 
	\includegraphics[width=\columnwidth]{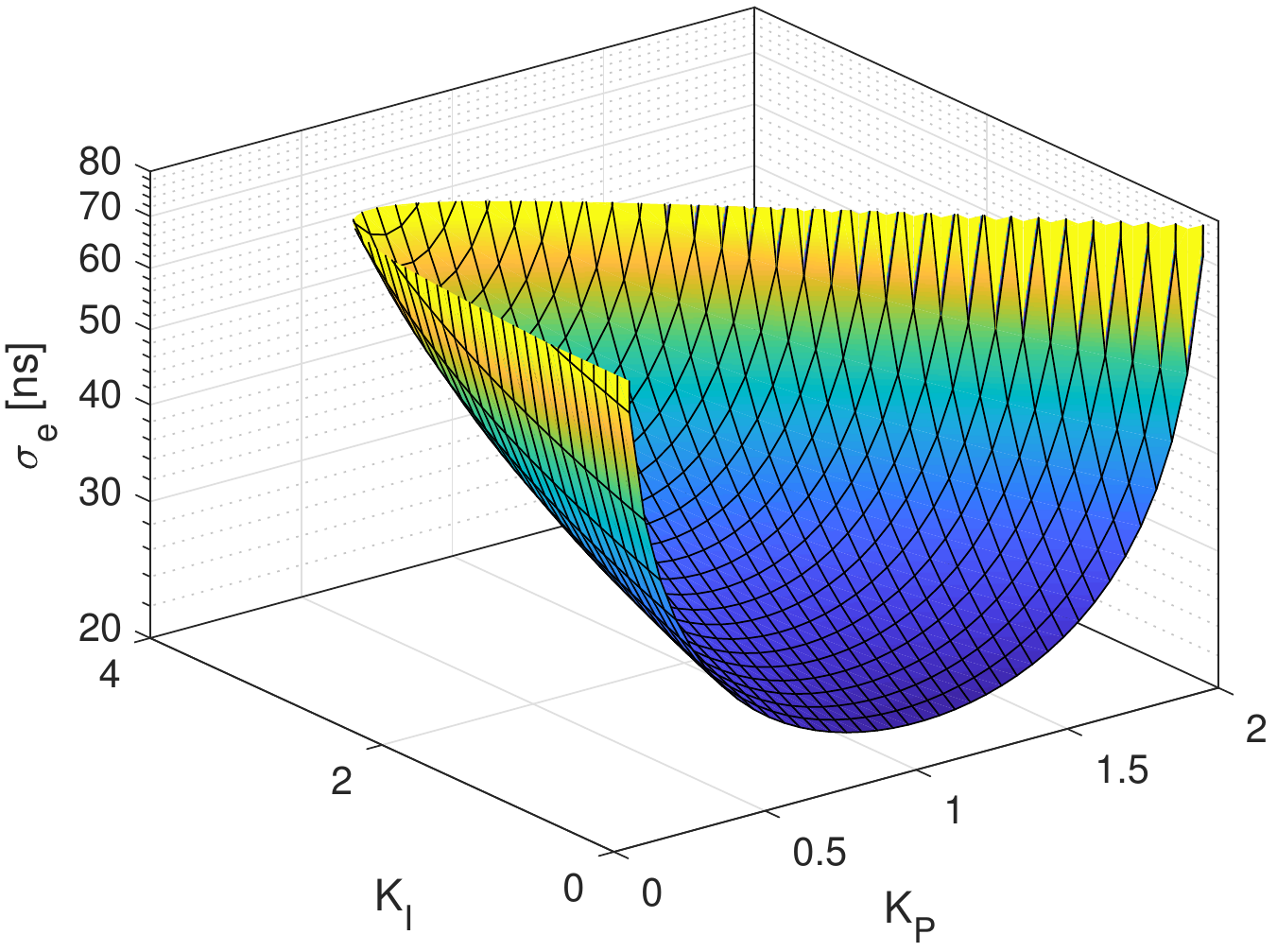}
	\caption{Standard deviation of the synchronization errors given by the square root of~\eqref{eq:Variance}, as a function of 
		the values of gains $K_P$ and $K_I$ in the stability region of the SC.
	}
	\label{fig:Ki_Kp}
\end{figure}
Since $q(k+1)$ is a random vector, its mean value is given by
\[
\E{q(k+1)} = A \E{q(k)} + G \E{\tau_{m_q}(k)} + Ff_0 ,
\]
where 
$\E{\cdot}$ denotes the expectation operator.  To compute
the uncertainty of $q(k+1)$ generated by the joint effect of $\nu(k)$
and $\eta(k)$, to a first approximation, we assume that: 
$\nu(k)\sim\mathcal{N}(0, \sigma_\nu^2)$, $\eta(k)\sim\mathcal{N}(0,
\sigma_\eta^2)$. 
As a result, the covariance matrix of $q(k+1)$ is given by
\begin{equation}
\label{eq:CovFirst}
 Q(k+1) = A Q(k) A^T + G G^T \sigma_\nu^2 + C C^T \sigma_\eta^2 .
\end{equation}
Of course, if $\nu(k)$ and $\eta(k)$ are not white or normally distributed,~\eqref{eq:CovFirst} holds just approximately. Notice that, by
setting $Q(0)$,~\eqref{eq:CovFirst} can be used to compute
the uncertainty of the state vector
in closed form.

If the controller gains are set so as to make~\eqref{eq:cltf}
stable,~\eqref{eq:CovFirst} reaches a steady-state equilibrium, i.e.,
there exists a sufficiently large value $\bar k$ such that $Q(k+1) =
Q(k)$, $\forall k > \bar k$.  Moreover, the steady-state variance of
the time error with respect to the reference 
is equivalent to 
the entry (1,1) of $Q(k)$. As a consequence, by computing the equilibrium of~\eqref{eq:CovFirst}, 
it follows that the synchronization error variance is 
\begin{equation}
\label{eq:Variance}
\sigma_e^2 = \E{(e(k) - \E{e(k)})^2} = \frac{2 (\sigma_\nu^2 + \sigma_\eta^2)}{K_P
	(4 - K_I - 2 K_P)} .
\end{equation}
Since $K_P > 0$, $K_I \geq 0$  and $\sigma_e^2 > 0$, it follows that the SC is stable for
$0 < K_P < 2$ and $K_I < 4 - 2 K_P$.
Within this region,~\eqref{eq:Variance}  is minimized once the
denominator is maximized. Therefore, the values of $K_P$ and $K_I$ minimizing the variance of the synchronization error can be determined from Fig.~\ref{fig:Ki_Kp}, which shows the behavior of  $\sigma_e$ as a function of $K_P$ and $K_I$ for values of $\sigma_\nu$ and $\sigma_\eta$ consistent with those of the system at hand and reported in Section~\ref{sec:results}. However, it is worth emphasizing that trend and position of the minimum do not depend on the variances of $\nu$ and $\eta$.

From this analysis it follows that if a prompt response is required, a
dead-beat controller, with $K_P = K_I = 1$, works well.  Otherwise, if
the uncertainty has to be minimized, $K_P = 1$ and $K_I = \varepsilon
> 0$ is the right choice, with $\varepsilon$ being a sufficiently small
constant. However, the smaller $K_I$, the longer the convergence time becomes.


\section{Experimental Results}
\label{sec:results}

Based on the previous analysis, the SC behavior was tested for different values of $K_P$ and $K_I$. 
The instrument used as a reference was a GPS-disciplined Meinberg~M600
master clock. 
The overall phase noise of the generated
PPS signal (estimated on the basis 
of the power spectral density reported in instrument's specifications) is about 30 ns. However, 
it is about one order of magnitude smaller over time 
intervals of a few hours, i.e. till when the
effect of flicker phase noise, white frequency noise, flicker frequency noise and random walk frequency noise are negligible. In such conditions, the phase noise is mainly white,
in accordance with the theoretical model described in Section~\ref{sec:model}.
The period fluctuations of the PPS signal generated by the SC in open-loop conditions
(namely when no control action is applied) were measured by an
Agilent DSO7032A with a 2-GHz sampling clock disciplined by the master clock.
The systematic
relative frequency offset of the emulated free-running DCO is about -82
ppm.
The standard deviation $\sigma_\eta$ of the
corresponding phase noise in open-loop conditions instead ranges from 
about 25 ns over one hour (when the phase noise is still dominated by white contributions) till
about 120 ns over two days (i.e. when the effect of the other low-frequency power-law noises becomes significant). Given that, as explained in Section~\ref{sec:model}, the models adopted for the SC design
rely on the inherent assumption that phase contributions are mainly white, $\sigma_\eta=25$ ns was also used to simulate
the behavior of the SC in closed loop for different values of $K_P$ and $K_I$. The resulting standard deviation values $\sigma_e$  are basically the same as those obtained from the square root of~\eqref{eq:Variance} and shown in 
Fig.~\ref{fig:Ki_Kp}. 

\begin{figure}[t]
	\centering
	\includegraphics[width=\columnwidth]{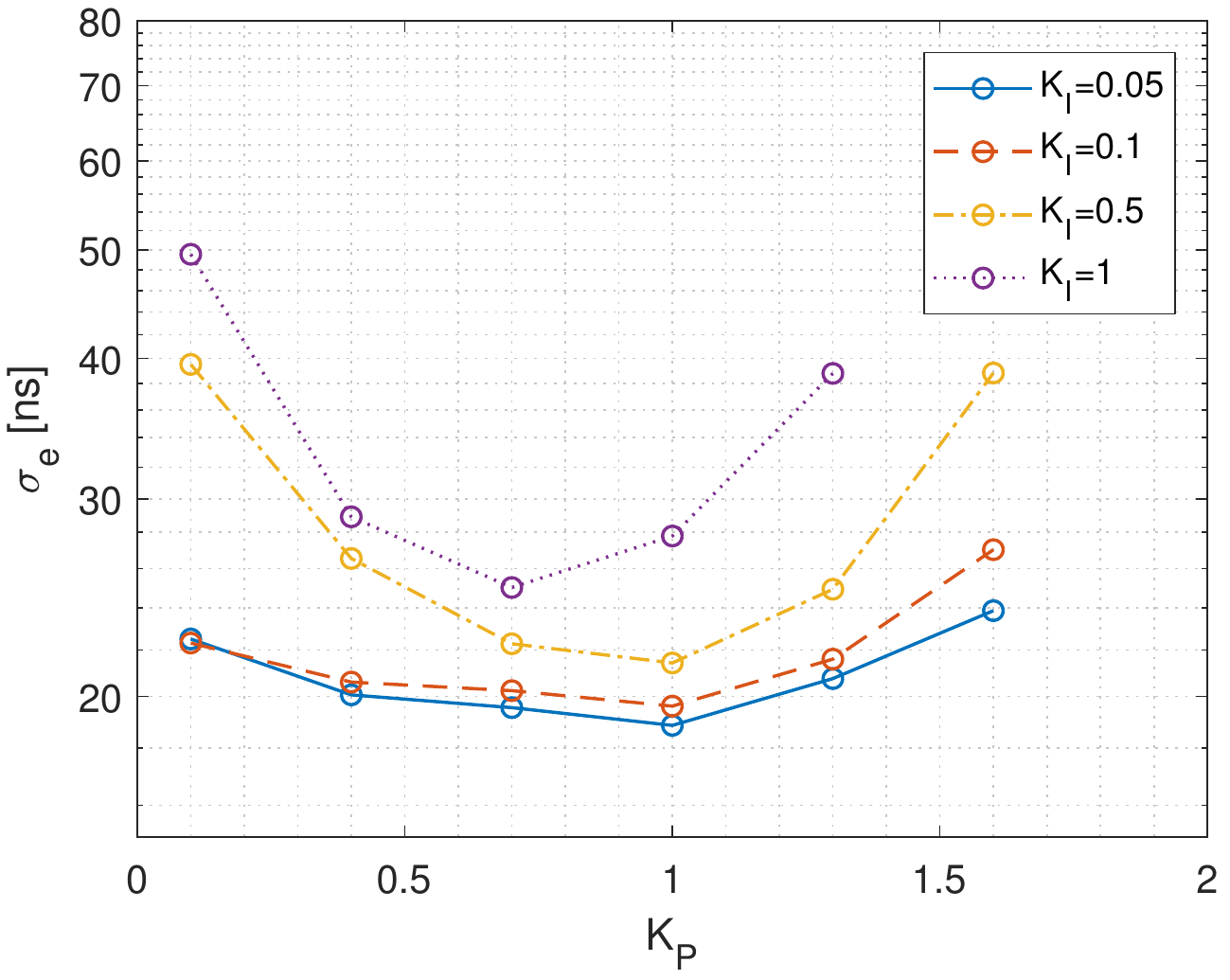}
	\caption{Experimental values of the standard deviation 
		of the	synchronization errors 
		for different settings of $K_P$ and $K_I$.}
	\label{fig:std_exp}
\end{figure}

The results of the theoretical analysis were also validated experimentally by estimating  $\sigma_e$
in steady-state conditions over 1-hour intervals for various pairs of $K_P$ and $K_I$ values within the SC stability region (i.e. with $K_I=\{0.05, 0.1, 0.5, 1\}$ and $K_P$ ranging
from $0.1$ and $1.6$). The corresponding standard deviation values are shown in Fig.~\ref{fig:std_exp}. 
The experimental curves clearly confirm that $\sigma_e$ exhibits a minimum value
when $K_P=1$ and  $K_I=\varepsilon>0$. However, for $K_I<0.05$ 
jitter reduction becomes negligible. Observe that even if experimental and theoretical results are generally quite consistent, significant deviations can be observed when both $K_P$ and $K_I$ tend to zero. 
The ultimate reason for this mismatch is not clear, but it is certainly
related to some second-order difference between SC model and SC implementation.
Nonetheless, it is worth noticing that the experimental results are smaller than those based 
on the theoretical analysis, which can be regarded as a conservative
design policy in this respect.
 
\begin{figure}[t]
  \centering
  \includegraphics[width=\columnwidth]{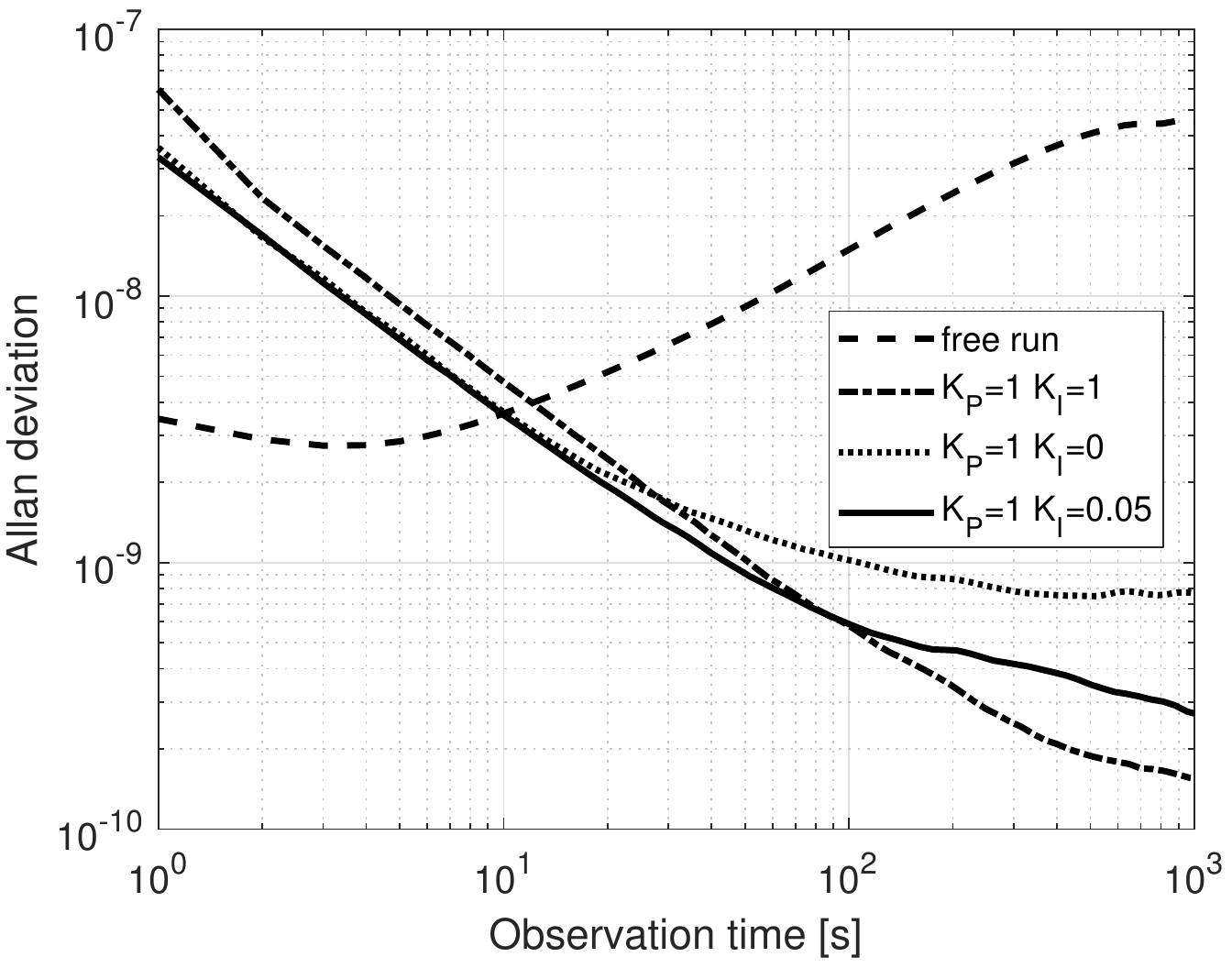}
  \caption{Allan deviation curves of PPS signals generated by the SC
    running on the BBB for different values of the PI controller
    parameters $K_P$ and $K_I$.  For the sake of comparison, also the
    Allan deviation of the PPS signal generated in open-loop
    conditions is shown. }
  \label{fig:allan}
\end{figure}

The long-term stability of the SC with respect to
the chosen reference PPS signal was determined by measuring the difference in
time (over about two days) between the rising edges of the PPS signals at the input and at
the output of the SC, respectively, for some of the $K_P$ and $K_I$ values considered in the design stage, i.e.
\begin{enumerate}
	\item Using a dead-beat controller (i.e. $K_P = K_I = 1$) that is
	chosen for its fast response, albeit the jitter reduction with this
	controller is not the best;
	\item Using a purely proportional controlled with unit gain (i.e. $K_P
	= 1$ and $K_I = 0$);
	\item And, finally, with $K_P = 1$ and $K_I = 0.05$ for the reasons explained above.
\end{enumerate}
The Allan
deviation values of the waveforms generated by the SC are shown in Fig.~\ref{fig:allan} for
different observation intervals.
The open-loop (i.e. free-running)
case is also shown for the sake of comparison. It is worth noticing
that the short-term stability of the  PPS in open-loop
conditions over 1 s is approximately $3.4\cdot10^{-9}$, but it tends
to degrade over longer intervals, as customary of low-quality crystal
oscillators.   Of course, in closed-loop conditions, the systematic relative
frequency offset with respect to the input reference signal is well
adjusted by the PI controller. Therefore, if the input reference
oscillator is particularly accurate, the systematic relative frequency
offset of the SC can be reduced to less than 0.1 ppm.  Observe that
the short-term SC stability is clearly worse than in open-loop
conditions, particularly when the dead-beat controller is used.
However, in the long term, stability with respect to the input
reference drastically improves as a result of the control action. 
Some test were performed also using different input PPS signal, 
i.e. the PPS from a u-blox NEO-6M. Such device produce a poorer 
quality PPS signal, hence in the long-term the performance is worse. 
Nevertheless it was possible to confirm the design choices 
discussed so far.
On the whole, the configuration based on the criterion described in
Section~\ref{sec:model} provides a very good trade-off between
short-term and long-term stability. 



\begin{figure}[t]
	\centering 
	\begin{tabular}{c}
		\includegraphics[width=0.97\columnwidth]{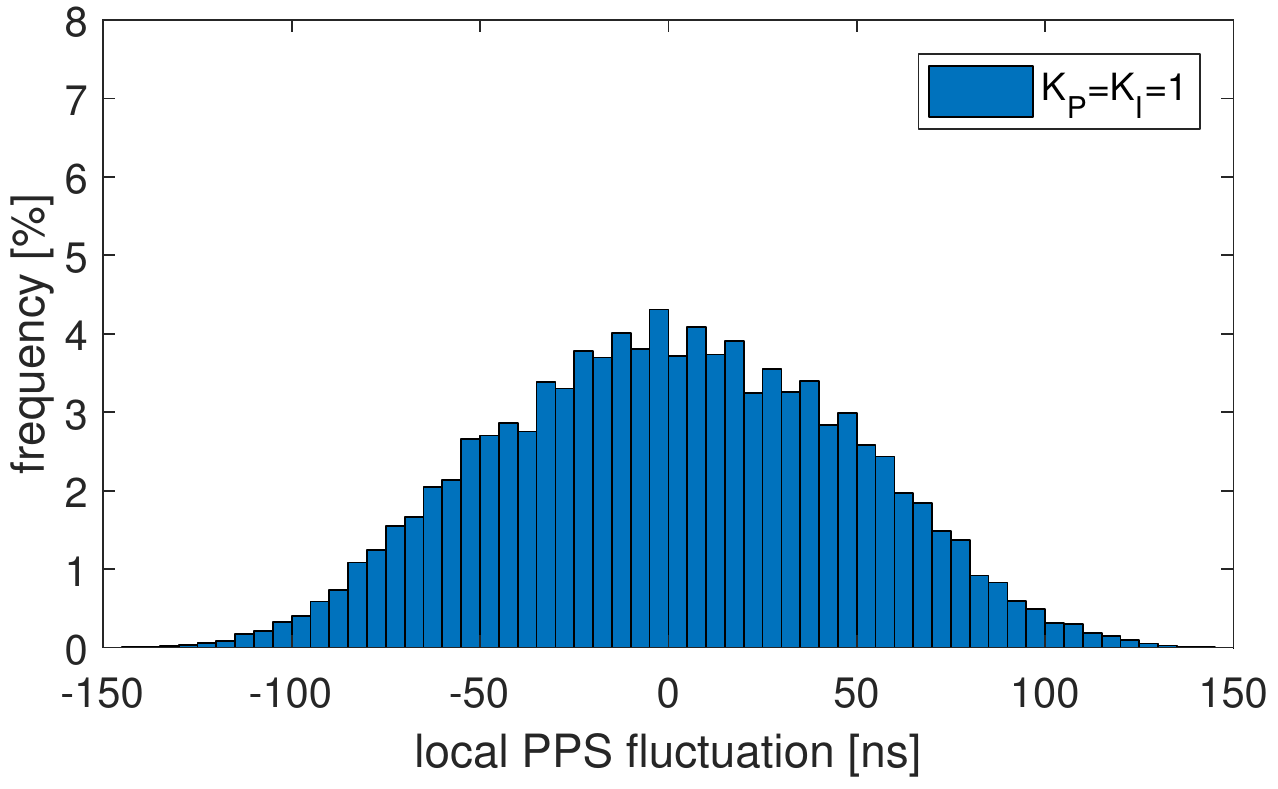} \\
		(a) \\
		\includegraphics[width=0.97\columnwidth]{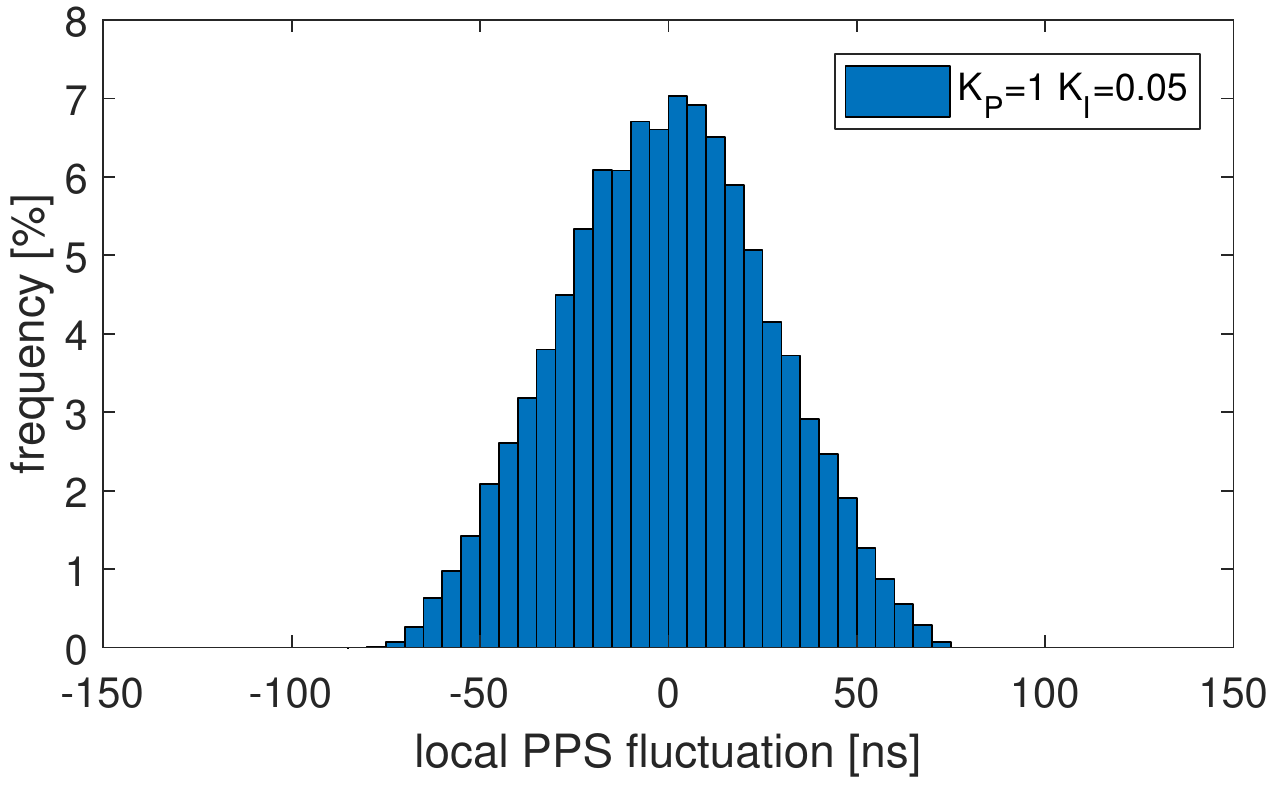} \\
		(b) \\
		\includegraphics[width=0.97\columnwidth]{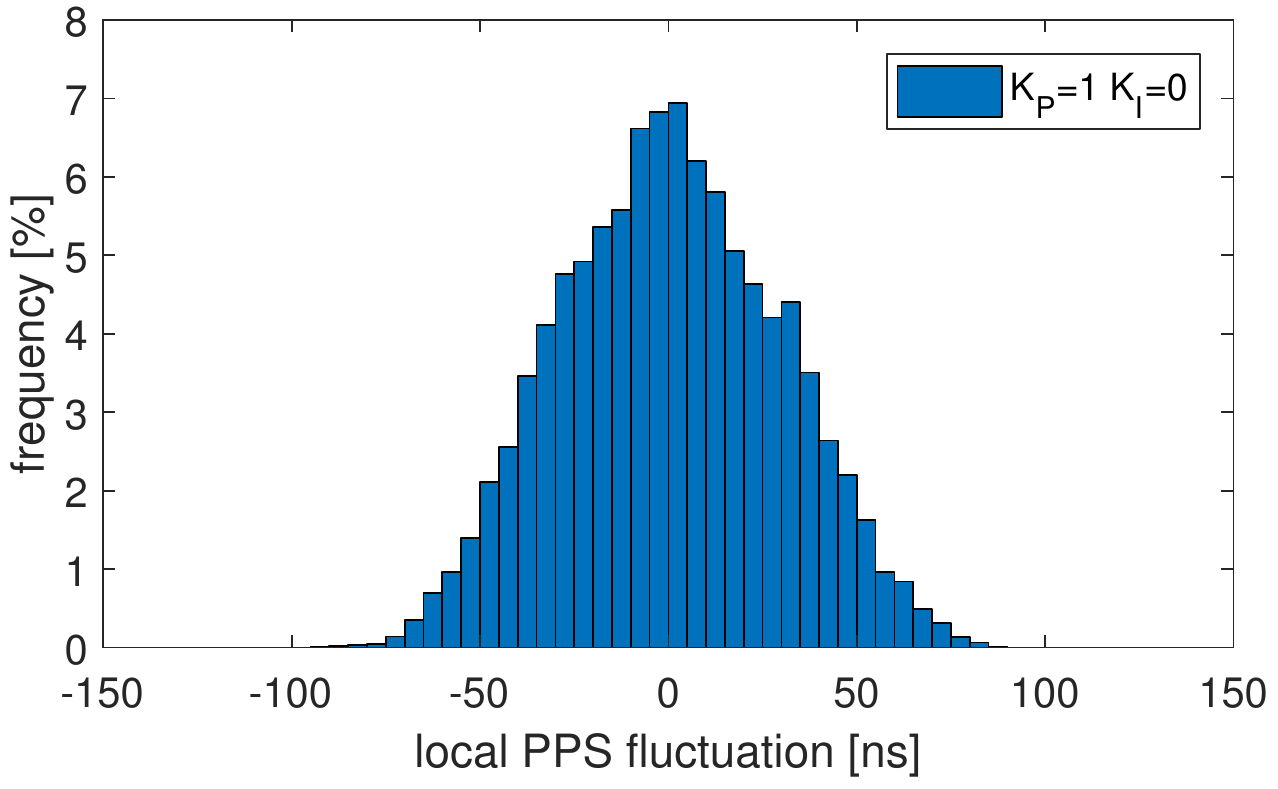} \\
		(c) \\
	\end{tabular}
	\caption{PPS period fluctuations of the signal generated by
          the SC with respect to the Meinberg~M600 using (a) a
          dead-beat controller ($K_P = 1$ and $K_I = 1$), (b) a
          quasi-optimal controller for time uncertainty minimization
          ($K_P = 1$ and $K_I = 0.05$) and (c) a proportional-only
          controller.  }
	\label{fig:histograms}
\end{figure}

In order to complete the analysis, Fig.~\ref{fig:histograms}(a)-(c)
shows the histograms of the PPS period fluctuations with respect to
the Meinberg~M600 clock using (a) the dead-beat controller; (b) the
quasi-optimal controller with $K_P =
1$ and $K_I = 0.05$ and (c) a purely proportional controller ($K_P =
1$ and $K_I = 0$).
Observe that in the second case, the jitter is almost halved (i.e. 25
ns vs. 50 ns).  The jitter associated wit the proportional-only controller 
is instead just slightly worse than the optimal one, whereas in theory it should be
much larger. Such a difference is due to the mismatch between theoretical model and
SC implementation explained before. In any case, the proportional-only controller
is not able to correct possible sudden time offsets perfectly.

\section{Conclusion}
\label{sec:conclusion}


This paper describes the design criteria of a
Servo Clock (SC) implemented on a single-board computer to discipline the data acquisition stage of a low-cost Phasor
Measurement Unit (PMU).
The SC has been developed in the context of the `OpenPMU' project.  The SC
implementation relies on a BeagleBone Black
(BBB) embedded platform. In particular, the SC runs mainly in one of the Programmable Real-time Units (PRUs) of
the BBB microprocessor, with no need for additional
hardware.
The PPS input reference signal could be provided by a common master clock shared among multiple PMUs within the same substation.
The experimental results are quite consistent with the theoretical ones and 
highlight the correct operation of the SC based on a PI controller correcting 
possible frequency offsets between the local oscillator of the BBB and the input reference.
The controller has been optimized in order to  minimize the standard deviation 
of the short-term synchronization error.
In the future, other and more sophisticated
(e.g. adaptive) techniques will be adopted to further optimize the
controller parameters on-line, e.g. to handle changes in environmental or processing load conditions.

\bibliographystyle{IEEEtran}
\bibliography{ispcs}

\end{document}